# Polymorphic self-poisoning in poly(lactic acid): a new phenomenon in polymer crystallization

Shu-Gui Yang[a,b], Xiang-bing Zeng[c,*], Feng Liu[a,b], Goran Ungar[a,c,*]

[a] State Key Laboratory for Mechanical Behavior of Materials, Shaanxi International Research Center for Soft Matter, School of Material Science and Engineering, Xi'an Jiaotong University, Xi'an 710049, China

[b] Institute of New Concept Sensors and Molecular Materials, Shaanxi Key Laboratory of New Conceptual Sensors and Molecular Materials, Xi'an Jiaotong University, Xi'an 710049, P. R. China

[c] School of Chemical, Materials and Biological Engineering, University of Sheffield, Sheffield S1 3JD, U.K.

ABSTRACT

Self-poisoning (SP) is ubiquitous in polymer crystallization, but has so far manifested itself visibly only as minima in growth rate *vs.* temperature in either monodisperse systems where e.g. unstable folded chains obstruct crystallization of stable extended chains, or in periodically segmented chains where unstable stems with n-1 segments disturb deposition of stable stems with n segments. Here we report a new type of self-poisoning found in poly(lactic acid) (PLA), where a less stable crystal form (alpha') disturbs growth of the stable form (alpha). While alpha requires strict up-down order of the polar chains, alpha' does not, hence is kinetically favoured. Unexpectedly, below the temperature of the growth rate minimum the lamellar thickness increases rather than drops, as in all other reported cases of polymer crystallization with decreasing temperature. A growth rate equation model is developed, giving good match with experiments, but revealing an unexpectedly low fold surface free energy of alpha' form. Due to SP of $\alpha$, most practical fast-cooling processing gives the low-modulus $\alpha$'-form grown close to $T_g$, explaining generally poor mechanical properties of the bio-friendly PLA.

INTRODUCTION

The self-poisoning (SP) phenomenon in crystallization of chain molecules was first recognized when it was found that crystal growth rate *G* of ultra-long monodisperse normal alkanes $C_nH_{2n+2}$ (120<*n*<390) reaches a maximum a few degrees below melting point $T_m^E$, then *decreases* to a sharp minimum near $T_m^F$, below which it increases again sharply [1]. Here $Tm^E$ and $Tm^F$ are melting points of crystals with chains fully extended (E) and folded in half (F), respectively [2]. This anomalous behavior is seen also in solution crystallization, where $T_m^i$ is replaced by dissolution temperature $T_d^j$ [3,4]. Below the minimum, growth of metastable but kinetically favoured F-crystals takes over from that of E-crystals. A second and third minimum was seen in longest alkanes on transitions to twice and trice-folded chain growth [5]. Moreover, G was also found to have a minimum as a function of increasing solution concentration c, with G dropping to zero at the E-F growth transition [4]. Conversely, starting from higher c, F growth stops as c is depleted, only to restart once an E nucleus forms causing local dilution that spreads through the solution as a dilution or "unpoisoning" wave, leaving E crystals behind [4,6].

SP was explained by the growth front being poisoned not by impurities but by native molecules themselves attaching in unstable "wrongly" folded conformations. These are just short of stability but are sufficiently long-lived to block the surface for productive growth of less folded or extended species [1,7,8]. Later a minimum was also observed in melt-crystallization of narrow fractions of ethyleneoxide oligomers [9].

The phenomenon exposed a fundamental limitation of the classical coarse-grain theory of polymer crystallization, which treats individual chain "stems" (straight traverses through lamellar crystals) attaching and detaching as whole units [10]. The theory could not reproduce the minima [11], whereas even a simple rate theory and MC simulation that split an extended chain into just two segments already achieved a semiquantitative match with experiment [4,7]. A more elaborate MC simulation was performed subsequently, highlighting the tortuosity of attachment of an extended chain, with a hiatus at half-length [12]. It has been pointed out that SP must also be hindering crystallization of polydisperse polymers, as the lingering "almost" sufficiently long stems block the progress of growth. Consequently, as shown by fine-grain simulations, lamellar growth faces are curved in the *xz* plane, where *x* and *z* are growth- and chain-axes [13,14].

In recent years Alamo et al reported multiple growth rate minima also in polydisperse polymers, but having substituent groups (halogen, ester) spaced at regular intervals along a polyethylene chain ("precision polymers") [15,16,17,18]. With decreasing crystallization temperature $T_c$, at each minimum lamellar thickness drops by the length of one monomer unit, e.g. from 4 to 3 and from 3 to 2 units. Our recent rate equation treatment, with stems split into monomer units, gave good quantitative fit to experimental growth-rate data for one such polymer series [19]. Based on simulations, Whitelam et al. outlined some key requirements for a system to show SP [20].

It should also be mentioned that, prompted by the discovery of SP in solution crystallization of alkanes, it has been suggested that previously unexplained inability of many proteins to grow crystals from solution past the size of nanoscale clusters is also due to SP [21]. The important role of SP in nucleation and growth of neurodegenerative amyloid buildup has been reported recently, and it was proposed that SP can be exploited to block amyloid formation [22].

Here we report a SP phenomenon in polymer crystallization different from those described previously. The competing, kinetically favoured growth taking over below the growth rate minimum is not producing thinner crystals, but a less ordered crystal form. In fact, surprisingly, it is giving thicker crystals, defying the normal rule that polymer lamellar thickness decreases with decreasing $T_c$. The material is the well-known mass-produced environmentally friendly poly(lactic acid) (PLA), and the two crystal forms in question are $\alpha$ and $\alpha$'. Below, after describing experimental results, we develop an analytical growth rate model of polymorphic SP, which reproduces well the essential features of the observations. The work adds new light to the currently debated role of mesophases in polymer crystallization, showing that a low-order alternative phase can hinder rather than assist crystal growth.

RESULTS AND DISCUSSION

*Experimental*

The left-handed PLA enantiomer, PLLA, of weight-average molecular weights $M_w$ (polydispersity) 9 kDa (1.2), 36 kDa (1.5) and 110 kDa (1.7), abbreviated PLLA-9k, PLLA-36k and PLLA-110k, were purchased from Jinan Daigang Biomaterial Co., Ltd. Their isothermal crystallization was studies first by DSC (TA DSC250), then by polarized optical microscopy (Olympus BX51-P with a Linkam LTS420E hot stage), by *in-situ* and *ex-situ* simultaneous small- and wide-angle X-ray scattering (SAXS/WAXS, Beamline BL16B1 at Shanghai Synchrotron Radiation Facility), and by scanning electron microscopy (SEM, ZEISS Sigma 300, etched with a water/methanol (1:2, v:v) solution containing 0.025 mol/L NaOH, then gold-decorated).

Linear growth rate of spherulites of the three polymers are plotted *vs* $T_c$ in Fig. 1. Between 121 and 106°C all three show a clear minimum, the most pronounced one in PLLA-9k. WAXS in Fig. 2 (bottom) shows that above the temperature of the minimum ($T_{min}$) the form is α (see also Fig. A1 in End Matter). The 110-200 d-spacing around 0.531 nm signifies the α-form, whereas that around 0.537 nm identifies the more loosely packed α'. Significantly (Fig. 2 bottom), there is an interval of ~10 K where the two forms crystallize simultaneously. This differs from previous SP cases, where the changeover between two morphologies was abrupt.

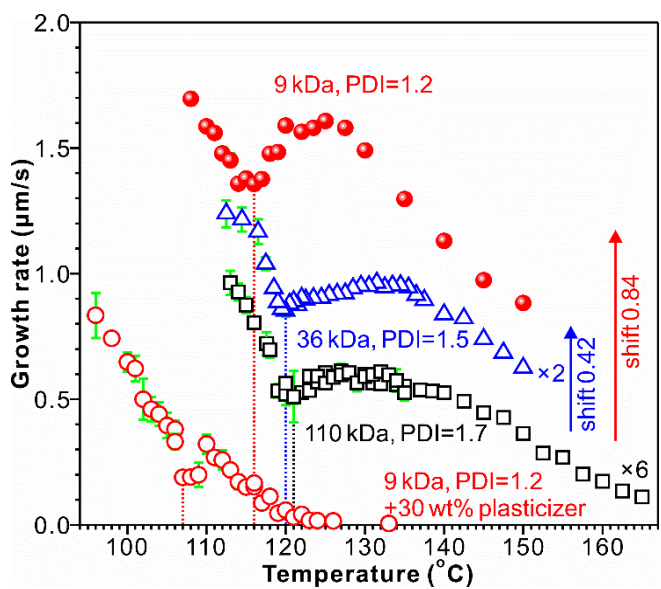

FIG. 1. Spherulite radial growth rate *vs* $T_c$ for PLLA-9k, PLLA-36k and PLLA-110k, and for PLLA-9k containing 30% plasticizer. Values for PLLA-36 and PLLA-110 were multiplied by 2 and 6; PLLA-36 and PLLA-9 datasets were shifted vertically as indicated.

A crystallization rate minimum between α and α' forms, a "discontinuity", has been noted several times before, observed by DSC [23,24] and direct growth rate measurement [25,26,27,28,29]. Some reports attributed the minimum to the Regime II – III growth transition [23,24,25,26], although with some reservations in [26]. However, regardless of one's opinion of growth regimes, they certainly cannot explain retardation with increasing supercooling. Other reports [25] describe the $G(T_c)$ curve as a double-bell, a single bell being a typical shape of $G(T_c)$ for a polymer between $T_m$ and glass transition $T_g$. This explanation attributes the retardation on the right of the minimum to increasing viscosity with lowering $T_c$. However, this fails to explain the sudden sharp growth acceleration to the left of $T_{min}$ [30]. Anyway, to test this interpretation we measured $G(T_c)$ in PLLA-9k containing 30% plasticizer methoxylated hydroxyethyl cardanol that suppressed $T_g$ from 45°C to 22°C. The resulting $G(T_c)$ is shown in Fig. 1 (empty circles) featuring a sharp minimum around 109°C that cannot be attributed to a thermally activated chain transport process. Incidentally, slowing chain transport had also been used in an attempt to explain the growth minimum in alkanes after it was first discovered [11].

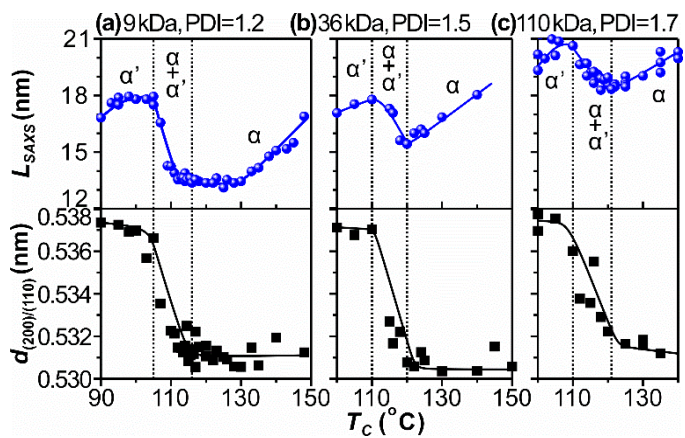

FIG. 2. X-ray data for the three polymers as a function of $T_c$. Top: SAXS long period; bottom: 110/200 WAXS lattice spacing. All the samples were crystallized isothermally in DSC apparatus, then quenched at room temperature (RT), recorded at RT. Raw SAX/WAXS data are at [31].

Most remarkably, the SAXS long spacing $L$ of PLLA *increases* rather than decreases as $T_c$ is lowered below $T_{min}$. This observation, unique in polymer crystallization [30], has already been made by Kawai et al [29] and Cho and Strobl [32], but they did not associate it with the change of form, as α' had not yet been identified. Using correlation function analysis, Cho and Strobl ascertained that the jump in $L$ was not merely due to a thicker amorphous layer, but to a genuinely increased crystal layer thickness $l_c$. In fact, our own determination of crystallinity $X$ by DSC shows that the increase in $l_c$ between α and α' is not only proportional to the increase in $L$, but even exceeds it. Using $\Delta H_m^0$ values for 100% crystalline α' and α forms [33], we obtain the following $X_{α'}$ ($X_α$ in brackets) for PLLA-9k, PLLA-36k and PLLA-110k: 0.69 (0.57), 0.68 (0.55) and 0.56 (0.51) (estimated error 0.03). This means that the relative increase in $l_c$ below $T_{min}$ is actually by 1/5 larger than suggested by the increase in $L$ in Fig. 2a.

It is noteworthy also that in the low-$M_w$ low-polydispersity PLLA-9k $L$ remains constant at ~13.5 nm over the entire 110-130°C $T_c$ interval. This is likely to be related to "integer folding" [2], i.e. to a preference of the chain making an integer number *i* of full crystal traverses. This leaves the chain ends at the crystal surface, thus reducing surface overcrowding [34,35]. The amorphous layer then takes the excess chain length plus missed traverses. Considering that the length of an average 10/3 helical PLLA-9k chain is 36 nm, in α and α'-forms close to $T_{min}$ *i* should be 5 and 3, respectively.

The micrograph in Fig. 3a shows an α spherulite grown at 125°C then cooled to 110°C. After 3 min, while the spherulite continues to grow, now in α'-form, many new α' nuclei appear (Fig. 3b). The SEM image in Fig. 3c shows the boundary between the α-spherulite and the α' grown epitaxially on it at 110°C.

For α' growth to take over from the more stable α-form at lower temperature, it must have a kinetic advantage. The main difference in structure of the two forms is that in α' there is up-down disorder in orientation of chains, the chain polarity being due to orientation of its ester groups [36]. By contrast, in α-form there is regular up-down chain alternation in {110} planes, and uniform up-up or down-down order in (200) planes [37]. Of the two forms, the less ordered α' has lower density, larger inter-chain distance and a significantly lower heat of fusion $\Delta H$ [33,38]. Unlike α, α' (or δ-form) belongs to the class of orientationally disordered, or plastic, crystals [39]. Its obvious kinetic advantage is that at a particular lattice site all chains can attach, while for α only half can do so, on average. Having an attachment rate twice that of α-stems, close to the temperature where wrongly oriented attachments are nearly stable and long-living, they would inhibit the growth of the stable α-form and act as poison.

In this respect, α' depositions act similarly to the way F-chains inhibit growth of E-crystals of alkanes, or 3-monomer long stems inhibit growth of 4-monomer thick lamellae of a "precision" polymer. However, there are significant differences in SP caused by polymorphism. Since heat of fusion of α' is considerably smaller than that of α ($\Delta H_\alpha \approx 1.4 \Delta H_{\alpha'}$) [33,38], its lamellae must grow thicker for the bulk crystallization energy to overcome the fold-surface free energy $\sigma_e$. This is reflected in the increase in *L* on transition from α to α' growth (Fig. 2). Thus, for the wrongly oriented stems to seriously inhibit the growth of α, they must be longer, creating a step-up in lamellar thickness (Fig. 3d,e). Once α' stem length exceeds the minimum number of repeat units $n_{min} = (2\sigma_e T_m^0)/(\Delta H \Delta T)$, α' growth can start taking over (here $T_m^0$ is equilibrium melting point and $\Delta T$ supercooling). However, in a limited *T*-range below $T_{min}$ α-growth still competes with α', since what it lacks in frequency of stem attachments it gains in their survival chance, which depends on $\Delta H$

ΔT. Hence the gradual increase in α' fraction and L within the ~10K interval below $T_{min}$ (Fig. 2). Below that interval α'-stem lifetime extends sufficiently for them to dominate the growth.

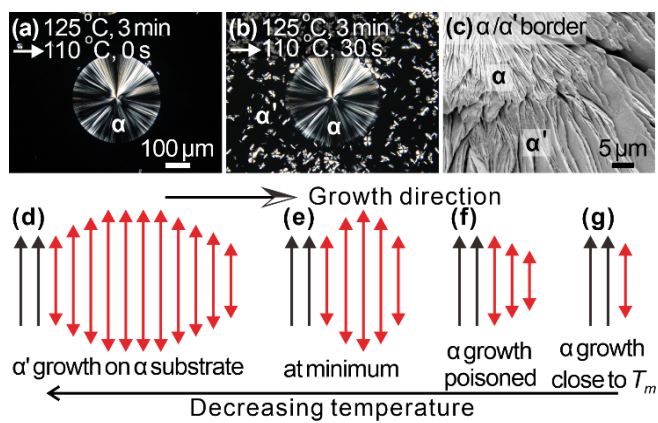

FIG. 3. (a-c) Micrographs of PLLA-9k; α form crystallized at 125°C for 3 min, then quickly cooled to 110°C to grow α'. (a,b) Polarized optical micrographs (a) immediately on reaching 110°C, and (b) after 30 s. (c) SEM of the same sample quenched from 110 °C. (d-g) Schematic "row-of-stems" model of growth of a PLA crystal lamella. Vertical single- and double-arrows represent stems of α-form, e.g. along a (020) plane, and of α'-form, respectively.

We propose that even well below $T_{min}$ ordered α pockets are expected to form where a sufficient number of stems happen to be "correctly" oriented in nearby crystallizing melt. Formation of such α pockets is supported by X-ray diffraction. Based on diffuse scattering, Tashiro et al. described their model of α'-form as a disordered conglomerate of local α-like domains [36].

It is remarkable that the presence of unstable attached overgrowth *thicker* than the stable growing parent lamella should be sufficient to block its growth, causing the rate minimum (Fig. 3e). The increase in thickness can be regarded as a process intermediate between secondary and primary nucleation. The high frequency of such nucleation is supported by the fact that many new α'-spherulites nucleate from melt as soon as $T_c$ is lowered to $T_{min}$ (Fig. 3b). Above $T_{min}$ new spherulites are scarce despite the 60-70 K supercooling of α, an indication that SP affects its primary nucleation probably even more than its growth.

*Theory*

We consider a "row of stems", normal to the crystal growth face, to set up the growth rate equation. The elementary steps are shown in Fig. 4a-d and described in the caption. The equations are developed in End Matter. Using these equations and the parameters in Table 1, the obtained growth rate is plotted in Fig. 4e.

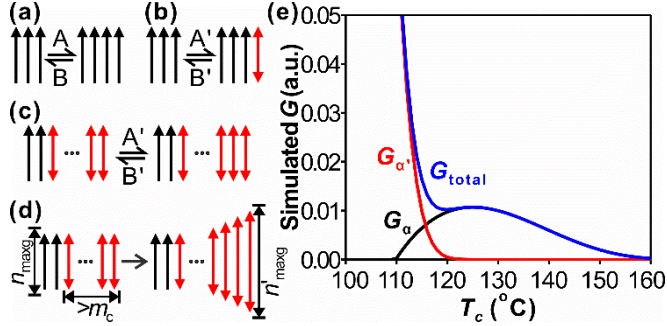

FIG. 4. (a-d) Schematic models used in simulation of measured growth rate. (a) Growth of $\alpha$-phase happens by attachment of a new $\alpha$-stem (black single arrow) to a clean unpoisoned growth front. The attachment rate is A and the detachment rate is B. (b) An $\alpha'$-stem (red double arrow) can attach to the $\alpha$-phase growth front, with attachment rate A' and detachment rate B'. (c) Poisoning of $\alpha$ growth as further $\alpha'$-stems can attach to the poisoned surface, but $\alpha$-stems can not. (d) $\alpha'$-phase can grow through thickening, but this is assumed to happen only when the number of $\alpha'$-stems at the poisoned growth front is over a critical value $m_c$. (e) Simulated growth rate. Orange: overall growth rate; blue and grey: growth rates of $\alpha$- and $\alpha'$-phases, respectively. The parameters used are given in Table 1.

Table 1. Parameters used to simulate the growth rate of PLLA.

| $\Delta H_\alpha$ (J/g) | $T_m^\alpha$ (°C) | $\sigma_e^\alpha$ (kJ/mol) | $K_1$ | $\Delta H_{\alpha'}$ (J/g) | $T_m^{\alpha'}$ (°C) | $\sigma_e^{\alpha'}$ (kJ/mol) | $K_3$ | $K_5$ |
|---|---|---|---|---|---|---|---|---|
| 104.5 | 185 | 14.5 | 0.1 | 71.6 | 165 | 8.95 | 0.1 | 1.5 |

In our calculation melting enthalpies $\Delta H_\alpha$ and $\Delta H_{\alpha'}$ were taken from literature [33]. The best-fit melting temperatures $T_m^\alpha$ and $T_m^{\alpha'}$ are within ~5K of those reported [40], and $\sigma_e^\alpha$ is also close to the experimental value of 15kJ/mol [41]. The model also holds an explanation of the notable fact that the supercooling at which $\alpha'$ starts crystallizing (165-120=45 K) is almost twice as large as that of $\alpha$ (185-160=25 K). Equation (14) in End Matter points to the nucleation process as the cause, i.e. to the fact that $\alpha'$ nuclei are more difficult to form because their lower crystallization energy $\Delta H_{\alpha'}$ requires larger lamellar thickness to overcome the end-surface free energy $2\sigma_e^{\alpha'}$.

In fact, since $\Delta H_{\alpha'} = \Delta H_\alpha/1.4$, one would expect a 50% increase in lamellar thickness as $\alpha'$ growth takes over from $\alpha$. This increase indeed happens in PLLA-9k, but it is nowhere near as large in the polydisperse 36k and 110k polymers (Fig. 2). The only way to understand the lower-than-expected jump in thickness in the latter polymers is to assume that $\sigma_e^{\alpha'} < \sigma_e^\alpha$, as shown by our simulation, where $\sigma_e^{\alpha'} = 0.6\sigma_e^\alpha$ (Table 1). To explain this difference in $\sigma_e$, we note that $\alpha'$-form contains a high degree of translational disorder along chain axis, along with a degree of conformational disorder. Notably, all observed sharp Bragg reflections of $\alpha'$ are either *hk*0 or 00*l*, with the (203) the only *hkl*-type [36]. Thus $\alpha'$ can almost be considered a mesophase. Previously one of us has indeed shown on the example of the real hexagonal columnar mesophase in 1,4-*trans*-polybutadiene that $\sigma_e^{meso}/\sigma_e^{crystal} \approx 0.4$ [42]. The case of the short-chain nearly monodisperse PLLA-9k, where $\sigma_e^{\alpha'} \approx \sigma_e^\alpha$, may be exceptional due to its integer chain-folding. These sizeable differences in $\sigma_e$ raise important general questions about the nature of fold-surface in polymers and the exact origin of $\sigma_e$.

CONCLUSIONS

A new type of self-poisoning has been identified that manifests itself as a minimum in crystal growth rate at the transition between $\alpha$- and $\alpha'$-form growth in PLA. Although the minimum has been observed previously, its origin had remained unexplained. Unlike in previous examples of SP, where growth of thinner lamellae inhibited the growth of thicker ones (in long alkanes and "precision" polymers), in PLA the competing lower-driving-force but lower-barrier process is the deposition of an orientationally-disordered crystal polymorph. Intriguingly, the competing $\alpha'$ crystallization that takes over below the rate minimum results in *thicker* lamellae than the high-*T* crystallization of the stable $\alpha$-form, a unique phenomenon in polymer crystallization. The rate-equation model developed gives reasonably close match with experimental kinetic data, considering its simplified "row-of-stems" nature. Due to SP of the $\alpha$-form, most fast-cooling processing gives the low-modulus $\alpha'$-form grown close to $T_g$, detrimental to mechanical properties of PLA. This most important of biodegradable polymers is also impaired by another type of SP, called "poisoning by purity; it hinders crystallization of the high-performance "stereocomplex" in a racemic mixture of PLA enantiomers, caused by local composition fluctuations being amplified, creating pockets of pure enantiomer blocking further stereocomplex growth [43]. The current findings of polymorphic SP have also broader implications since $G(T_c)$ minima, or "bimodal" rate curves, have been observed in a number of key commercial polymers (polypropylene [44], Nylons [45,46,47]) at high supercooling using fast chip calorimetry. While alternative explanations have been proposed, polymorphic SP is now an open possibility to be considered.

DATA AVAILABILITY

Raw SAXS/WAXS data are freely available at [31].

ACKNOWLEDGEMENTS

We acknowledge financial support from the National Natural Science Foundation of China (52373022, 52003215, 22250710137), the Engineering and Physical Science Research Council UK (EP-T003294). and Shaanxi Provincial Science and Technology Department (2025GH-YBXM-042). The authors thank the staff at BL16B1 beamline at Shanghai Synchrotron Radiation Facility for help with X-ray experiments.

END MATTER

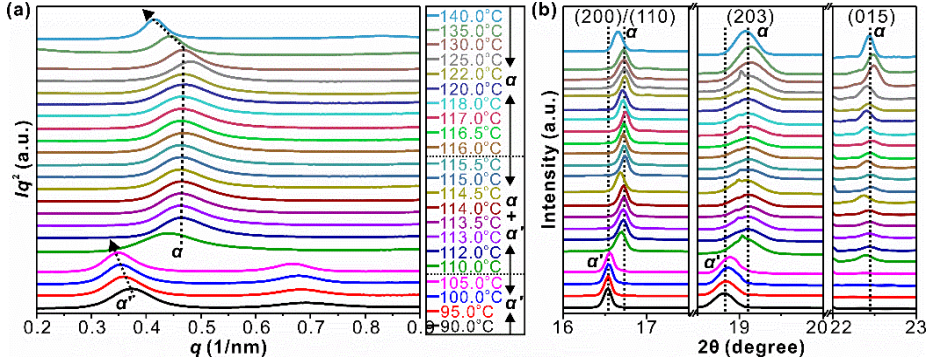

**Fig. A1. a**) and (**b**) are *ex-situ* SAXS and WAXS profiles of PLLA-9k isothermally crystallized at different temperatures.

## *Growth rate equation and its derivation*

For $\alpha$ form, without poisoning by the $\alpha'$ form, growth rate $G_\alpha = A - B = A(1 - \frac{B}{A})$. Here $A$ is attachment rate of stems on the growth front, also regarded as the barrier factor, while $1 - B/A$ is the driving force factor [13]. The latter is linked to the free energy difference between crystal and melt

$$\frac{B}{A} = \exp\left(-\frac{n\Delta T \Delta S - 2\sigma_e}{kT}\right) \quad (1)$$

Here $\Delta T = T_m^\alpha - T$, $\Delta S$ is the entropy loss at crystallisation per repeat unit, $n$ is the number of repeat units in the crystalline stem, and $2\sigma_e$ the surface energy per chain. The minimum length of the crystallized stem, as given in the main text, can be derived using the condition $A = B$ or $\frac{B}{A} = 1$ (Fig. 4a).

We assume the free energy barrier $F_B$ for stem attachment is entropic, therefore proportional to $T$ and the number of repeat units $n$, so

$$F_B = KnT \quad (2)$$

Here $K$ is a constant. $F_B$ can be taken as the free energy barrier of a repeat unit to get ready to attach to the growth surface, corrected by the fraction of the whole stem that needs to attach to the surface first in order for the rest of the stem to follow. Therefore

$$A = A_0 \exp(-\frac{KnT}{kT}) = A_0 \exp(-K_1 n) \quad (3)$$

where $K_1 = \frac{K}{k}$. Combining equations (1) and (4) we have

$$B = A_0 \exp(-K_1 n) \exp\left(-\frac{\Delta T \Delta S}{kT} n\right) \exp(\frac{2\sigma_e}{kT}) = A_0 \exp\left(\frac{2\sigma_e}{kT}\right) \exp[-(K_1 + K_2)n] \quad (4)$$

where $\frac{\Delta T \Delta S}{kT} = K_2$ and

$$A - B = A_0 \exp(-K_1 n) - A_0 \exp\left(\frac{2\sigma_e}{kT}\right) \exp[-(K_1 + K_2)n] \quad (5)$$

Therefore without poisoning, on the basis of equation (5), the maximum growth rate can be found at

$$n_{mg} = [\frac{2\sigma_e}{kT} + \ln(1 + K_2/K_1)]/K_2 = n_{min} + \ln(1 + K_2/K_1)/K_2 \quad (6)$$

The growth rate is then

$$G_\alpha = A_0 \frac{K_2}{K_1 + K_2} \exp(-K_1 n_{mg}) \quad (7)$$

For $\alpha'$ form, similarly we define

$K_3 = \frac{K'}{k}$ and $K_4 = \frac{\Delta T' \Delta S'}{kT}$. Maximum growth rate is found at

$$n'_{mg} = [\frac{2\sigma'_e}{kT} + \ln(1 + K_4/K_3)]/K_4 = n'_{min} + \ln(1 + K_4/K_3)/K_4 \quad (8)$$

and growth rate of $\alpha'$ form

$$G_{\alpha'} = A'_0 \frac{K_4}{K_3 + K_4} \exp(-K_3 n'_{mg}) \quad (9)$$

In our simulation we have assumed $A'_0 = 2A_0$ (that the $\alpha'$ stems can attach to the growth front either up or down), and $K_3 = K_1$.

When $\alpha$ form of length $n$ is poisoned by the unstable $\alpha'$ form, $\alpha$ form can only grow on the fraction $f_\alpha$ of growth surface that is not covered by $\alpha'$ chains (Fig. 4a-c), given by $f_\alpha = \left(1 - \frac{A'}{B'}\right)$, where $A' < B'$ are the respective attachment and detachment rates of an $\alpha'$ chain on the growth surface. So

$$G_\alpha = (A - B)(1 - \frac{A'}{B'}) \quad (10)$$

Here, $\frac{A'}{B'} = \exp\left(\frac{n \Delta T' \Delta S' - 2\sigma'_e}{kT}\right) = \exp(-\frac{2\sigma'_e}{kT}) \exp(K_4 n)$.

The growth rate with poisoning is therefore

$$G_\alpha = A_0 \exp(-K_1 n) - A_0 \exp\left(\frac{2\sigma_e}{kT}\right) \exp[-(K_1 + K_2)n] - A_0 \exp\left(-\frac{2\sigma_e}{kT}\right) \exp[-(K_1 - K_4)n] + A_0 \exp[-(K_1 + K_2 - K_4)n] \quad (11)$$

We can assume that the poisoning does not change $n_{mg}$ determined by equation [8]. Then

$$G_\alpha = A_0 \frac{K_2}{K_1+K_2} \exp(-K_1 n_{mg}) \left[1 - \exp\left(-\frac{2\sigma'_e}{kT}\right)\exp(K_4 n_{mg})\right] \qquad (12)$$

At the same time, we consider the possible growth rate of $\alpha'$ form through surface mediated extension of $\alpha'$ stems at the growth front. We assume that such extension happens only when the number of attached $\alpha'$ stems at the growth front is larger than a critical value $m_c$ (Fig 4d), which is considered to be proportional to the difference between $n'_{mg}$ and $n_{mg}$

$$m_c = K_5(n'_{mg} - n_{mg}) \qquad (13)$$

The fraction $f_{\alpha'}$ of the growth front with number of $\alpha'$ stems higher than $n_c$ is

$$f_{\alpha'} = \left(\frac{A'}{B'}\right)^{K_5(n'_{mg}-n_{mg})} \qquad (14)$$

Now the growth rate of $\alpha'$ is

$$G_{\alpha'} = A'_0 \frac{K_4}{K_3+K_4} \exp(-K_3 n'_{mg}) \left(\frac{A'}{B'}\right)^{K_5(n'_{mg}-n_{mg})} =$$
$$A'_0 \frac{K_4}{K_3+K_4} \exp(-K_3 n'_{mg}) \exp\left(K_5(n'_{mg} - n_{mg})\frac{n\Delta T' \Delta S' - 2\sigma'_e}{kT}\right) \qquad (15)$$

The overall growth rate is

$$G = G_\alpha + G_{\alpha'} \qquad (16)$$

15